\documentclass{article}
\usepackage{aem}
\usepackage{amssymb}
\usepackage{amsmath}
\usepackage{amsfonts,mathrsfs,amsthm}
\usepackage[pdftex]{graphicx}  % Add graphics capabilities
\usepackage{float}
\newcommand{\av}[1] {\left\langle #1 \right\rangle}
\newcommand{\modulc}[1]{\left\lvert #1 \right\lvert ^{2}}
\newcommand{\modul}[1]{\left\lvert #1 \right\lvert }

\newcommand{\Imm}[1]{\textrm{Im}\left [ #1 \right ]}
\newcommand{\Ree}[1]{\textrm{Re} \left [ #1 \right ]}
\newcommand{\En}{\vec{\cal{E}}_{n} }

\title{Statistics of the electromagnetic response of a chaotic reverberation chamber}

\name{J.-B. Gros$^{1*}$, U. Kuhl$^1$, O. Legrand$^1$, F. Mortessagne$^1$, O. Picon$^2$, E. Richalot$^2$}

\address{$^1$Universit\'e Nice-Sophia Antipolis, CNRS, Laboratoire Physique de la Mati\`ere Condens\'ee,\\
UMR 7336, 06100 Nice, France, \\
$^2$Universit\'e Paris-Est, ESYCOM (EA 2552), UPEMLV, ESIEE-Paris, CNAM, 77454 Marne-la-Vall\'ee, France \\
*corresponding author, E-mail: {\tt jean-baptiste.gros@unice.fr}
}

\begin{document}
\maketitle

\begin{abstract}
\normalsize{This article presents a study of the electromagnetic response of a chaotic reverberation chamber (RC) in the presence of losses.
By means of simulations and of experiments, the fluctuations in the maxima of the field obtained in a conventional mode-stirred RC are compared with those in a chaotic RC in the neighborhood of the Lowest Useable Frequency (LUF). The present work illustrates that the universal spectral and spatial statistical properties of chaotic RCs allow to meet more adequately the criteria required by the Standard IEC 61000-4-21 to perform tests of electromagnetic compatibility.}
\end{abstract}

\section{Introduction}

The electromagnetic (EM) reverberation chambers (RC) are commonly used for electromagnetic compatibility tests. Thanks to the presence of a mechanical stirrer and losses leading to modal overlap, the systems under test are submitted to a supposedly statistically isotropic, uniform and depolarized electromagnetic field \cite {Standard}. These properties are generally well realized if the excitation frequency is much larger than \emph{the lowest useable frequency} (LUF). It is important to underline here that the LUF depends strongly at the same time on the size of the chamber and on the importance of the modal overlap induced by the losses (either related to Ohmic dissipation at walls, to junction imperfections, or to antennas) \cite {Cozza}. The statistical behavior of the field, required in an RC above the LUF, is supposed to rely on the validity of Hill's hypothesis, according to which the EM field can be considered as a random superposition of plane waves \cite {Hill_1, Hill_2}. It is generally admitted that this hypothesis is realized for efficient stirring conditions.
Yet it turns out that this statistical behavior is naturally realized for most resonant modes in a chaotic cavity without resorting to any stirring \cite { GROS_2013}. Therefore one can take advantage of the universal statistical properties of the chaotic cavities to improve the behavior of the RC in the neighborhood of the LUF. It is what we wish to demonstrate here by comparing the EM responses in a chaotic RC and in a commercial mode-stirred RC, obtained via a numerical model and experimental measurements.

\section{How to make a commercial RC chaotic}

\begin{figure}
\centerline{
\includegraphics[width=0.5\columnwidth]{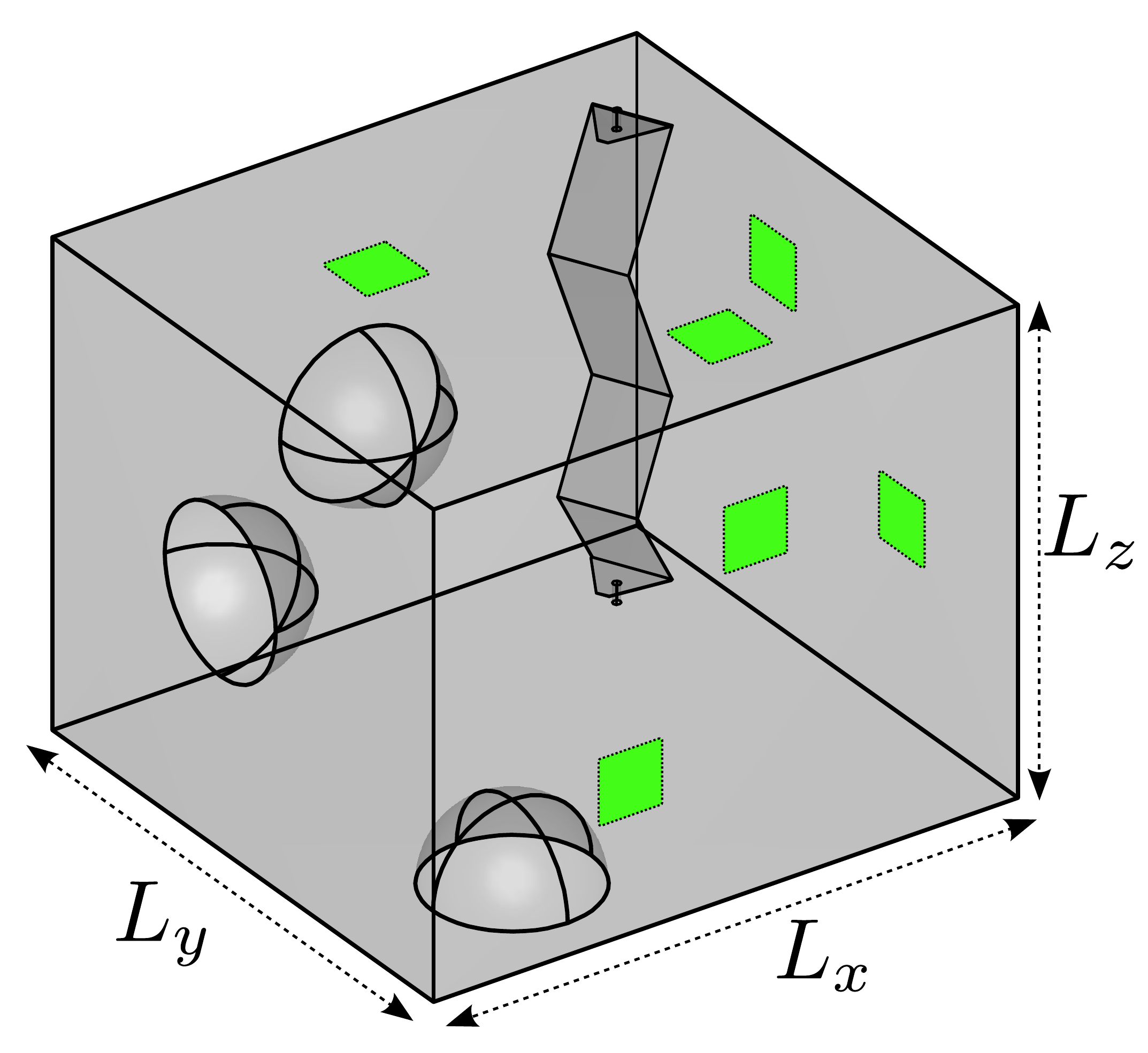}
\includegraphics[width=0.5\columnwidth]{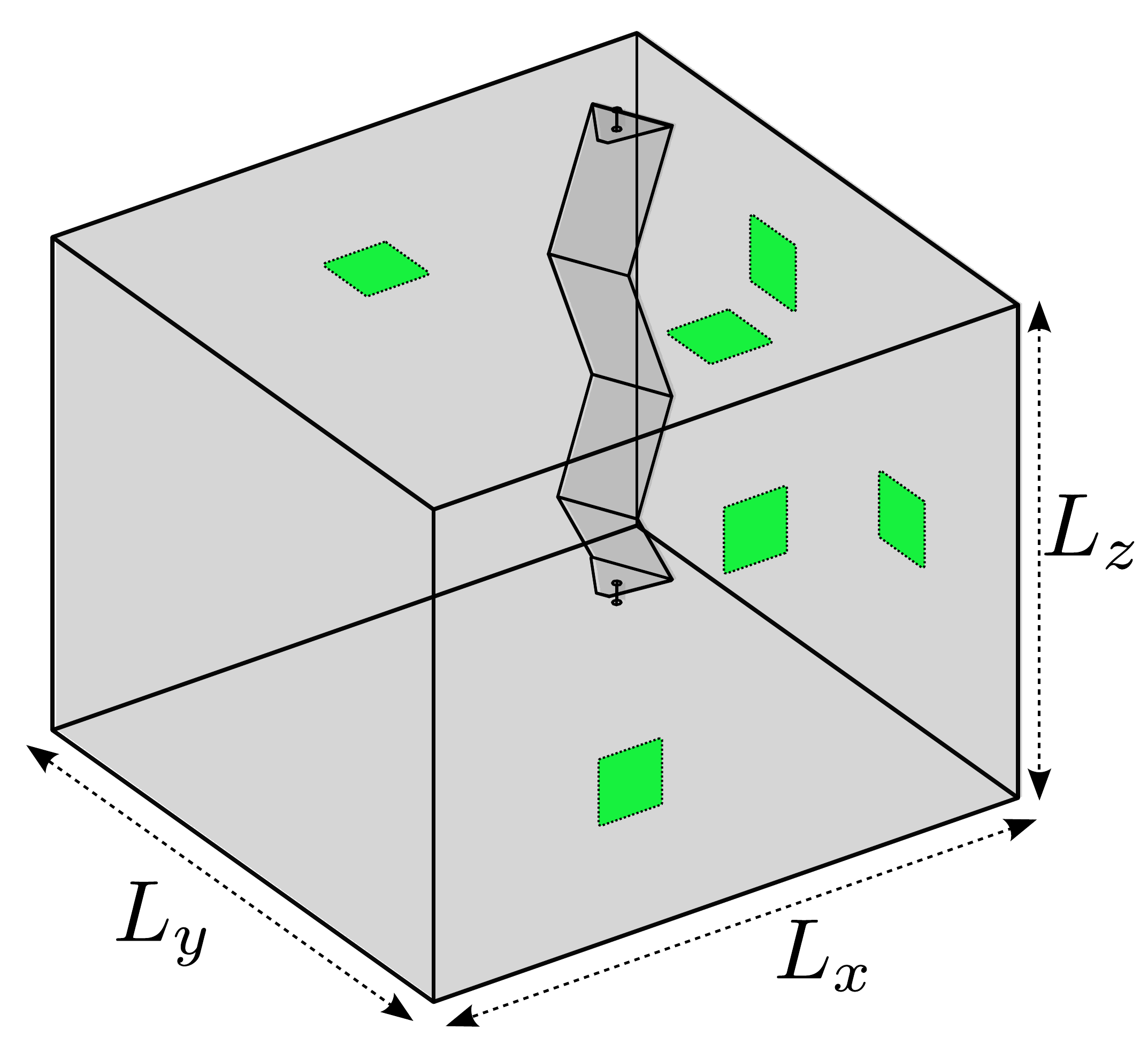}}
\caption{\label{fig:cav}
Left: RC made chaotic through the introduction of 3 half-spheres. Right: Conventional RC. The small squares correspond to Ohmic absorptive patches. For both cavities, the dimensions are: $L_x=2.951$ m, $L_y=2.751$ m, $L_z=2.354$ m.}
\end{figure}

\begin{figure}[t]
\centerline{
\includegraphics[width=\columnwidth]{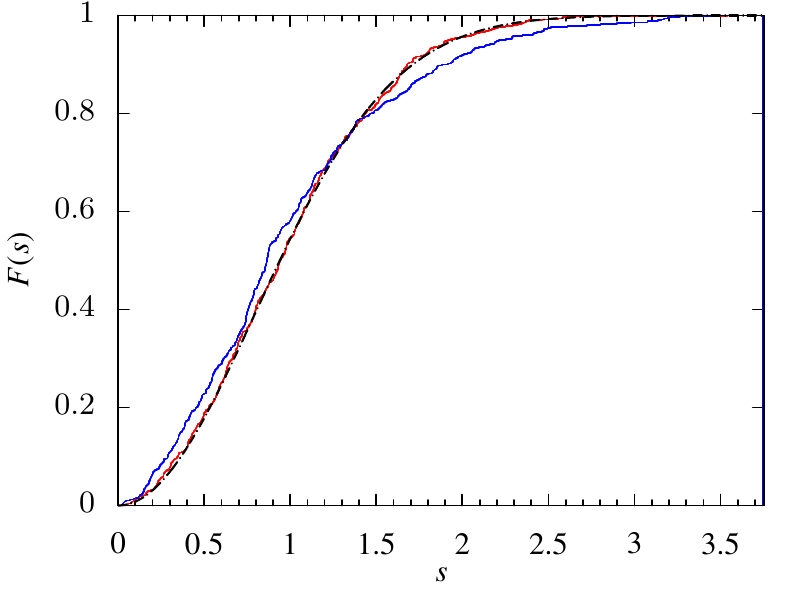}}
\caption{\label{fig:pde_s}
CDFs of the normalized eigenfrequency spacings for the conventional RC (blue curve) and for the chaotic RC (red curve) are compared to the Wigner Surmise (dash-dotted).}
\end{figure}

Here we study a chaotic RC (left of Fig.~\ref{fig:cav}) and a conventional RC equipped with a stirrer (right of Fig.~\ref{fig:cav}).
To assess the chaoticity of the cavity shown in the left side of Fig.~\ref{fig:cav}, we use the methodology presented in \cite{GROS_2013} which proposes to use the predictions of the random matrix theory (RMT) as quantitative criteria. Indeed, since the Bohigas-Giannoni-Schmit conjecture \cite{BGS} concerning the universality of level fluctuations in chaotic quantum spectra, it has become customary to analyse spectral statistics of chaotic cavities with the help of statistical tools introduced by RMT. To this end, in both RCs of Fig.~\ref{fig:cav}, we computed 880 eigenmodes of the lossless cavities for a fixed position of the stirrer and deduced the cumulated distribution function (CDF) of the normalized frequency spacings.
Those CDFs are compared in Fig.~\ref{fig:pde_s} with the corresponding CDF of the Wigner surmise which turns out to be an excellent approximation of the exact result deduced from the Gaussian orthogonal ensemble (GOE) of RMT, known to predict the universal statistical behavior of chaotic cavities. Only the RC with the hemispheres complies with the GOE prediction (Fig.~\ref{fig:pde_s}). It is a clear demonstration that the stirrer itself is not adequate to make the RC fully chaotic. As demonstrated in \cite{GROS_2013}, by reducing drastically the amount of parallel walls, the introduction of the three half-spheres is an efficient way to suppress almost all regular modes which are still present in the conventional RC.

\section{ Modelling the response of an RC in the presence of losses}

Now, we model the influence of Ohmic losses on the EM response inside an RC, at frequency $f$, via the Dyadic Green's tensor (DGT) with complex values:
\begin{equation}
 \overline{\overline {G}}(\vec{r},\vec{r}_0,f)=\sum_{n=1}^{\infty} \frac{k_n^2 \vec{E}_n(\vec{r})\otimes\vec{E}_n(\vec{r}_0)} {k^2(k_n^2-k^2)} \label{eq:Green}
\end{equation}
where the sum runs over the complex resonances defined by their eigenvalues $k_n=\frac{2\pi f_n}{c}(1-\frac{i}{2 Q_n})$ with $f_n$ the central frequency of the $n$th resonance and $f_n/Q_n=\Gamma_n$ its width, and the complex eigenfields $\vec{E}_n$ at the measurement point $\vec{r}$ and at the excitation point $\vec{r}_0$ (a current pointlike source).
In the literature, the DGT is often written as a matrix:
\begin{equation}\label{eq:FGD_mat}
\overline{\overline {G}}=\left(
\begin{matrix}
  G_{xx} & G_{xy} & G_{xz}\\
  G_{yx} & G_{yy} & G_{yz}\\
  G_{zx} & G_{zy} & G_{zz}
\end{matrix}
\right)
\end{equation}
where each column ($i=x$, $y$, ou $z$) contains the three Cartesian components of the electric vector field for an excitation polarized along $\vec{e}_i$.

In the case of a chaotic RC, the complexness parameter of modes, defined by:
\begin{equation}
q^2_n=\frac{\av{\Imm{\En}\cdot\Imm{\En}}}{\av{\Ree{\En}\cdot\Ree{\En}}}
\end{equation}
(where the vectorial field is appropriately normalized via the transformation $\En \rightarrow \vec{E}_n/\sqrt{\iiint\vec{E}_n\cdot\vec{E}_n dv}$ which cancels the global phase of each component \cite{PRGKLM}) as well the widths $\Gamma_n$ of resonances, verify the statistical predictions obtained with the help of RMT applied to open chaotic systems (see for instance \cite{KLM} and references therein and \cite{GROS_2014} for a more recent theoretical and experimental investigation of width shift distribution in the chaotic RC discussed in section 5 of the present paper).

In the studied range of frequencies, for the dimensions indicated in Fig.~\ref{fig:cav}, if losses were only due to the finite conductivity of the walls, the mean quality factor would be of the order of $10^4-10^5$. Still, in practice, the latter is rather of the order of a few $10^3$ due to losses related to antennas, imperfections at junctions and to sundry objects introduced in the RC. For the sake of simplicity, in our models, the losses are introduced through identical \emph{patches} (Fig.~\ref{fig:cav}), distributed over the walls, with a conductivity chosen to ensure a mean quality factor ($\av{Q_n}$) of the RC ranging from 1500 to 2000 (which are realistic values in RCs with such dimensions around 400 MHz). Perfect metallic boundaries were imposed on other parts of the walls. The complex resonances ($k_n$ and $\En$) in both RCs of Fig.~\ref{fig:cav} were obtained via a FEM software.

\section{Statistical study}
\subsection{Statistics of the elements of the DGT}

In order to perform a statistical study, we consider 30 different configurations for each RC, which were obtained by rotating the stirrer around its axis. For each configuration, the DGT Equation~(\ref{eq:Green}) was numerically computed at 300 frequencies of excitation regularly spaced between 7$f_c$ and 7.3$f_c $ (where $f_c$ is the cut-off frequency of one of the configurations of the chaotic RC). This interval, situated in the neighborhood of the 370th mode, roughly consists of 50 resonances whose mean modal overlap, defined by $d=\av{\Gamma_n}/\Delta f=8 \pi V f^3c^{-3}\av{Q_n}^{-1}$ (where $\Delta f $ is the mean spacing between adjacent resonant frequencies and $V$ is the volume of the RC), is approximately equal to 0.45.

\begin{figure}
\centerline{
\includegraphics[width=\columnwidth]{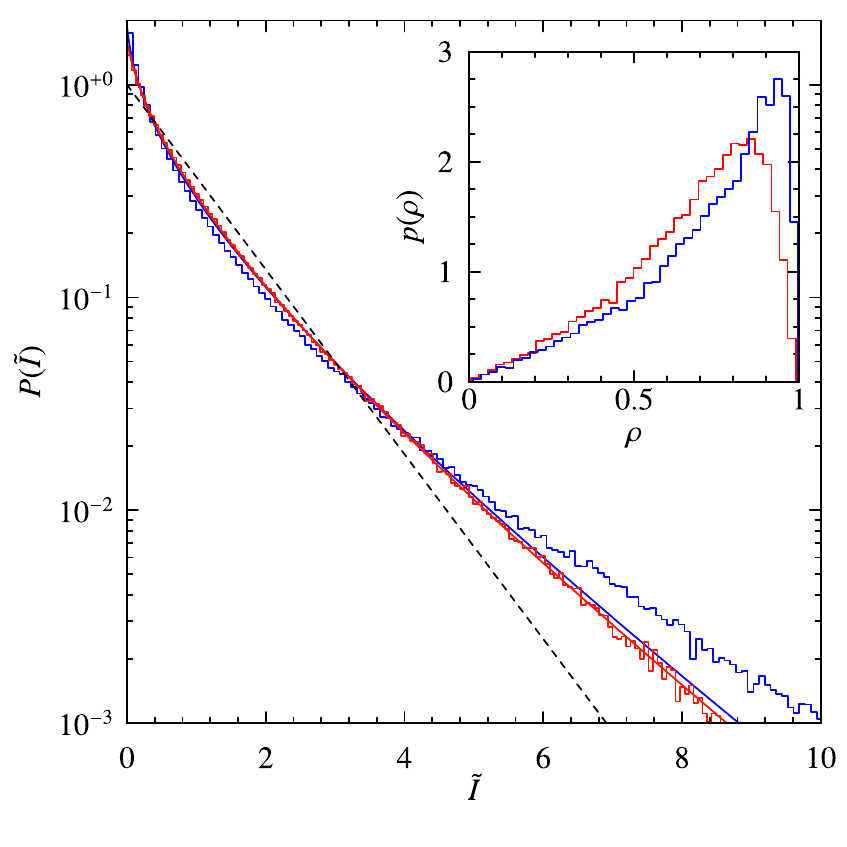}}
\caption{\label{fig:Pde:I:et:rho}
Empirical distributions of the normalised intensities: blue histogram for the conventional RC and red histogram for the chaotic one. Comparison with the prediction Equation~(\ref{eq:P:de:I:int}) (blue and red continuous curves) where the empirical $p(\rho)$ was used (see inset with corresponding colors). The dashed black curve shows the exponential distribution expected under Hill's hypothesis}
\end{figure}

For frequencies much larger than the LUF, where in fact $d\gg 1$, Hill's hypothesis is generally verified. The latter leads to a complex EM field, each Cartesian component of which has real and imaginary parts which are statistically independent and identically distributed following a normal distribution. In this case, the distribution of the squared modulus of each component follows an exponential law. This regime has been extensively explored in other contexts such as in nuclear physics (Ericson's regime) and in room acoustics (Schroeder's regime) \cite{ERICSCHRO}.
Yet, in the regime we are actually concerned with ($d \lesssim 1$), the real and imaginary parts of each component of the field are not identically distributed \cite{GROS_2013,PRGKLM}. For a given frequency of excitation and a given configuration, in the case of an ideally chaotic RC, they still are distributed according to normal laws, but with different variances. The ensuing distribution of the squared modulus of each component $I_i=\modulc{E_i}$ is then no longer exponential and depends on a single parameter $\rho_i$, called the \emph{phase rigidity}, defined by:
\begin{equation}
\rho_i=\frac{\av{E_i^2}}{\av{\modulc{E_i}}}\; .
\label{def_rho}
\end{equation}
More precisely, the distribution of $\tilde{I_i}=I_i/\av{I_i}$ depends on the sole modulus of $\rho$ through the expression \cite{KIM}
\begin{equation}\label{P_de_I}
  P_{\rho_i}(\tilde{I_i})=\frac{1}{\sqrt{1-\modulc{\rho_i}}} \exp\left[ -\frac{\tilde{I_i}}{1-\modulc{\rho_i}} \right] I_o\left[\frac{\modul{\rho_i}\tilde{I_i}}{1-\modulc{\rho_i}} \right]
\end{equation}
This distribution has been previously proposed by Pnini and Shapiro in \cite{Pnini} in order to model partially open chaotic systems and provides an interpolation between the two ideal distributions, namely Porter-Thomas for closed systems ($\modul{\rho} \rightarrow 1$) and exponential for completely open systems ($\modul{\rho}=0$ ). Recently, in \cite{Arnaut}, this distribution was compared to empirical distributions obtained in a conventional RC at low frequency. Unfortunately, such a comparison did not take into account the non-universal character of the conventional RCs at low frequency, nor even the fact that $\rho$ is a parameter which is distributed with frequency as well as with configuration \cite{KIM} as shown in the inset of Fig.~\ref{fig:Pde:I:et:rho}. Indeed, this parameter has to fluctuate either through a mechanical or electronic stirring. In the chaotic RC, due to the isotropy of the field, we observed that, for a given frequency of excitation and a given configuration, each columnn of the DGT is associated to a single value of $\rho$.
Thus, through any type of stirring (mechanical or electronic), the resulting distribution of normalised intensities of any Cartesian component is given by:
\begin{equation}\label{eq:P:de:I:int}
  P(\tilde{I})=\int_0^1 P_\rho(\tilde{I}) p(\rho) d\rho
\end{equation}
This result is illustrated in Fig.~\ref{fig:Pde:I:et:rho} where the empirical distributions of the normalised intensities of all the components (for all the excitation frequencies and all configurations) for both RCs are compared with the above equation, where the empirical distribution of $\rho$ was used. The excellent agreement between the prediction (\ref{eq:P:de:I:int}) and the empirical distribution associated to the chaotic RC (red curve and histogram) clearly demonstrates that the assumptions of statistical uniformity and isotropy (from which Equations (\ref{P_de_I}) and (\ref{eq:P:de:I:int}) are deduced) are only verified in the chaotic RC. One should note that the blue histogram related to the conventional RC agrees neither with Equation~(\ref{eq:P:de:I:int}) (blue curve) nor with the exponential distribution expected when Hill's hypothesis holds (dashed black curve).

\subsection{Field uniformity criterion from the International Standard}

\begin{figure}
\centerline{
\includegraphics[width=\columnwidth]{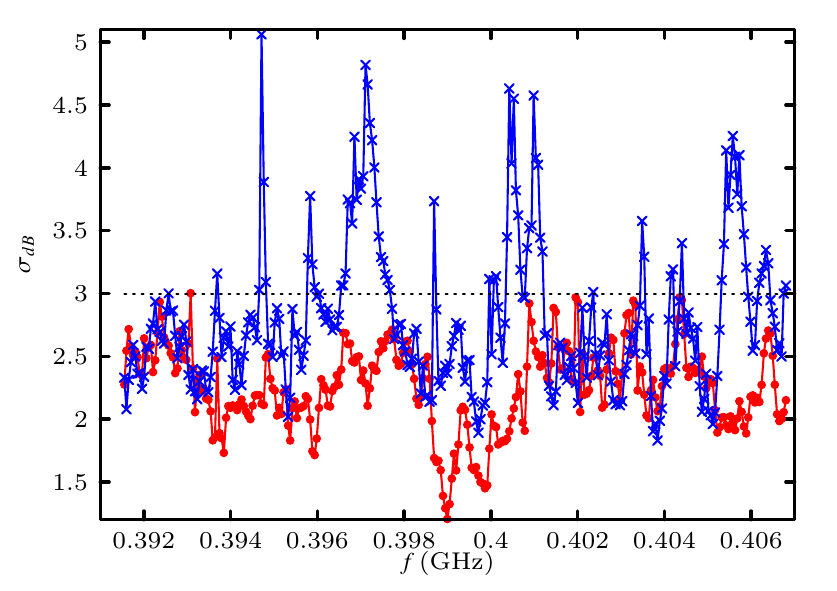}
}
\caption{\label{fig:fluctu}
Fluctuations of maxima in dB. Chaotic RC: red connected points. Conventional RC: blue connected crosses.}
\end{figure}

Here we are interested in the fluctuations of the maxima of the field amplitude evaluated through \cite{Standard}:
\begin{equation}\label{fluc}
\sigma_{dB}(f)=20 \log_{10}\left(1+\frac{\sigma_{\textrm{max}}}{\langle\modul{E_a}{}_{\textrm{max}}\rangle}\right)
\end{equation}
According to \cite{Standard}, one chooses 8 measurement points (distant from at least a quarter of a wavelength). At each of theses points, for 30 uncorrelated positions of the stirrer, one computes the DGT (\ref{eq:FGD_mat}) (each columnest being divided by the square root of the input power) and one keeps, for each line, the component with maximum modulus $\modul{E_a}{}_{\textrm{max}}=\modul{\max \left( G_{ai}\right)}{}$. One then computes the average and the standard deviation over the 8$\times$3 values of
$\modul{E_a}{}_{\textrm{max}}=\left\lbrace \modul{E_x}{}_{\textrm{max}},\modul{E_y}{}_{\textrm{max}},\modul{E_z}{}_{\textrm{max}}\right\rbrace$.
In the International Standard \cite{Standard}, the field is assumed to be uniform when $\sigma_{dB}<3$dB. The chaotic RC (red curve of Fig.~\ref{fig:fluctu}) complies almost always with this criterion and in a much better way than the conventional RC does (blue curve of Fig.~\ref{fig:fluctu}).

\begin{figure}
\centerline{
\includegraphics[width=\columnwidth]{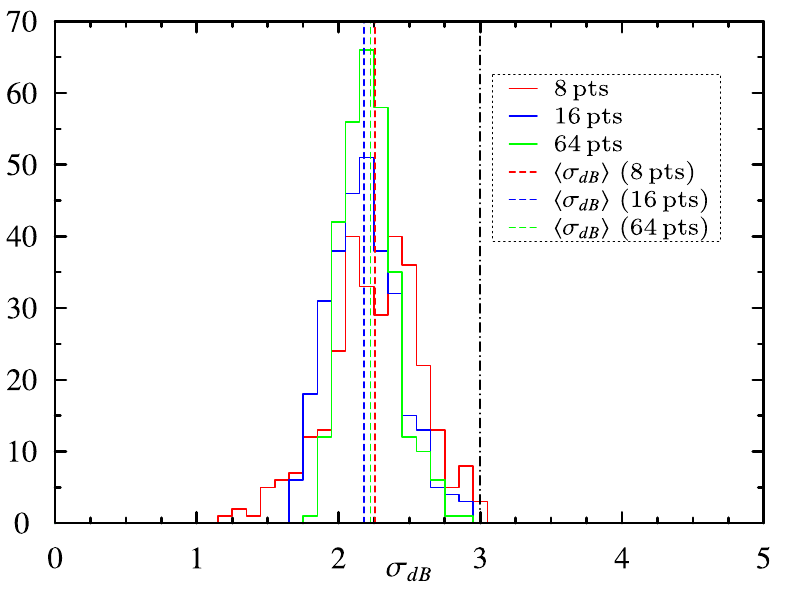}
}
\caption{\label{fig:hist_chaos}
Histograms of values of $\sigma_{dB}$ obtained for 8 (red), 16 (blue), 64 (green) measurement points in the chaotic RC. The red, blue and green dashes represent respectively the average value of $\sigma_{dB}$ for 8, 16 and 64 points. The limit at 3 dB is also shown. The percentages of values above this limit are respectively: 0.33\,$\%$, 0\,$\%$ and 0\,$\%$.}
\end{figure}

\begin{figure}
\centerline{
\includegraphics[width=\columnwidth]{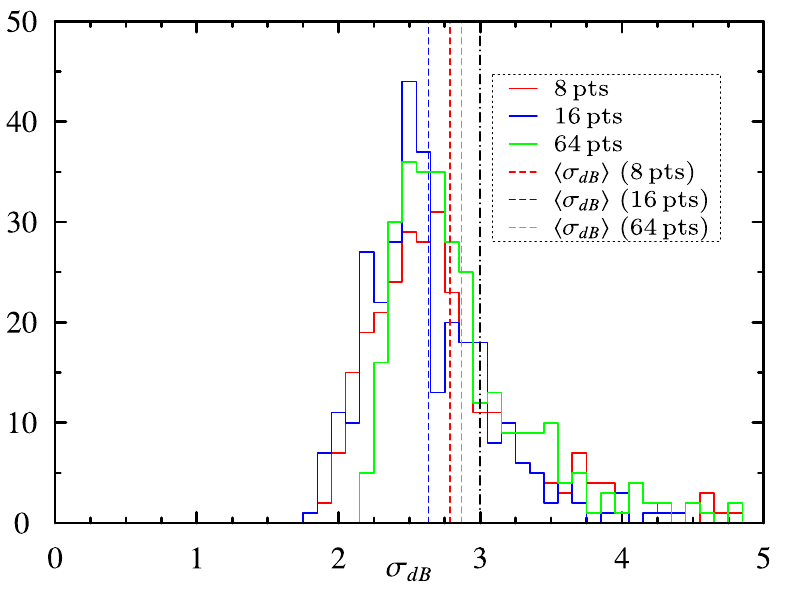}
}
\caption{\label{fig:hist_bras}
Histograms of values of $\sigma_{dB}$ obtained for 8 (red), 16 (blue), 64 (green) measurement points in the conventional RC. The red, blue and green dashes represent respectively the average value of $\sigma_{dB}$ for 8, 16 and 64 points. The limit at 3 dB is also shown. The percentages of values above this limit are respectively: 25.7\,$\%$,16\,$\%$ et 27.3\,$\%$.}
\end{figure}

In order to quantify more adequately these different behaviors, we propose to represent them through histograms and to consider how they change as the number of measurement points is varied (see Fig.~\ref{fig:hist_chaos} and Fig.~\ref{fig:hist_bras}).
The histograms shown in Fig.~\ref{fig:hist_chaos} correspond respectively to 8, 16 and 64 measurement points in the chaotic RC. It is clear that increasing the number of points does not modify the average significantly (red, blue and green dashes superimposed) and tends to concentrate the histograms around a unique average value which is definitely less than 3\,dB.

In complete contradistinction, the histograms presented in Fig.~\ref{fig:hist_bras} associated to the conventional RC, exhibit a behavior depending on the number of points. Indeed, the average values significantly depend on the number of measurement points and are much higher than in the chaotic RC. Moreover, the dispersions do not tend to be reduced when the number of points increases. This clearly demonstrates the non universality of the field statistics obtained in a non-chaotic RC.

On the basis of the numerical results presented above, it is more than plausible that the universal statistical features of the response in a chaotic RC guarantees a more important reliability of the criterion of uniformity proposed by the Standard. The strong sensibility of the criterion of uniformity towards the number of measurement points in a conventional RC, in the neighborhood of the LUF, tends to demonstrate its inadequacy in a non-chaotic RC.

\section{Experimental study}

\begin{figure}[!h]
\centerline{
\includegraphics[width=7cm]{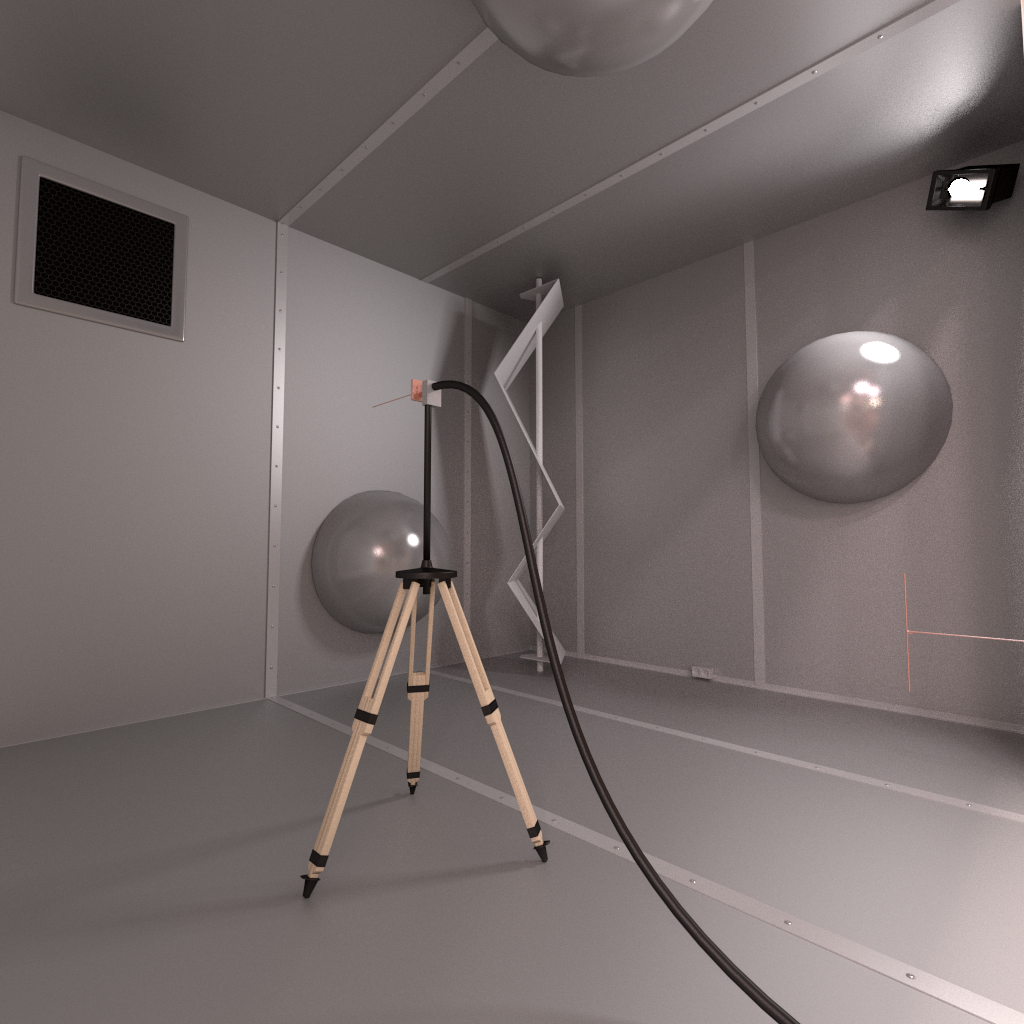}
}
\caption{\label{fig:manip_schem}
Artist 3D-view of the reverberation chamber made chaotic through the addition of 3 half-spheres. The volume of the RC is 19.1\,m$^3$ without the half-spheres, each having a radius of 40\,cm.}
\end{figure}

\begin{figure}
  \centerline{\includegraphics[width=\columnwidth]{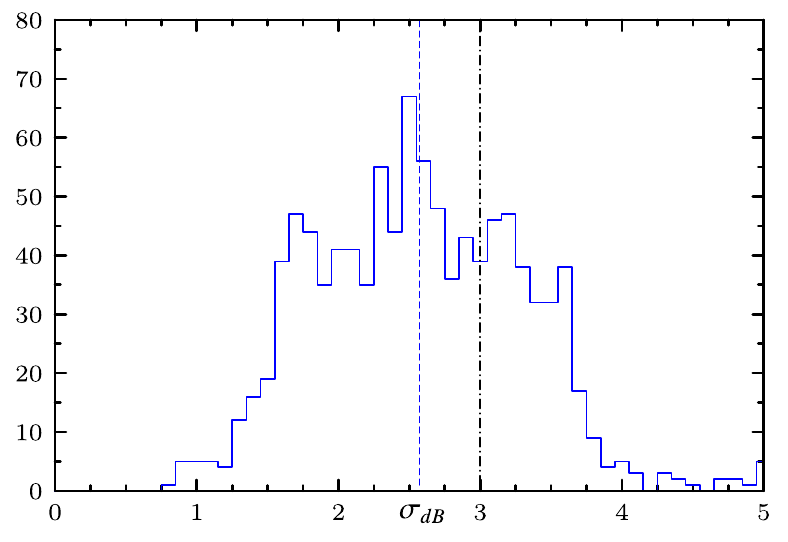}}
  \centerline{\includegraphics[width=\columnwidth]{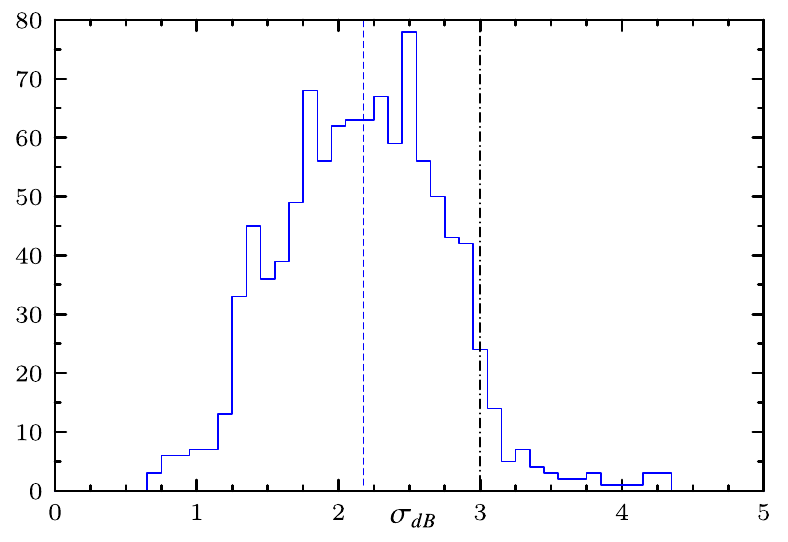}}
\caption{\label{fig:res_exp}
Histograms of values of $\sigma_{dB}$ experimentally obtained in the two configurations of the RC shown in Fig.~\ref{fig:manip_schem}.
Also shown: the limit at 3dB (black dash-dots) and the mean of $\sigma_{dB}$ (blue dashes).
Above: histogram for the conventional RC with the stirrer ($\av{\sigma_{dB}}=2.57$). Below: histogram for the chaotic RC with 3 half-spheres and the stirrer ($\av{\sigma_{dB}}=2.18$).}
\end{figure}

This section presents experiments that were performed in a commercial RC equipped with a vertical stirrer as already mentioned above.
This RC is made chaotic by the addition of 3 metallic half-spheres on its walls (cf Fig.~\ref{fig:manip_schem}). The chaotic character of this configuration was verified by the methods described in the reference \cite{GROS_2013}.
In both configurations of this RC (bare or with half-spheres) ($V\!\simeq$ 20 m$^3$), the $S$-matrix was measured between two antennas (one dipole and one monopole). Measurements were realized for 1024 regularly spaced frequencies in a frequency range of 20\,MHz centered around 400\,MHz, for 30 positions of the stirrer spaced by 12 degrees and for 8 different positions of the monopole antenna inside the volume. After extracting the coupling strength of the antennas \cite{KLM,PRGKLM}, one can deduce from the measurement of $S_{12}$ the normalized value of the amplitude of the Cartesian component of the field along the monopole antenna whose orientation is kept fixed. The mean quality factor is estimated to be around 2000 and appears to be almost insensitive to the introduction of the half-spheres. In the frequency range of our study (of the order of 5 to 6 times the cut-off frequency), the modal overlap $d\!\sim$\! 0.45 remains moderate. $d$ is deduced directly from the measurements by extracting the complex resonances using the method of harmonic inversion\cite{GROS_2014}.
From the measured responses, we will concentrate on the analysis of the criterion of uniformity. As in the previous section, in Fig.~\ref{fig:res_exp} we present a comparison of histograms of $\sigma_{dB}$ obtained experimentally in the two configurations mentioned above. These preliminary measurements tend to confirm the numerical results presented in the previous section. Indeed, the criterion of uniformity is verified in a much better way in the chaotic configuration, in particular because of a reduced dispersal of the distribution of $\sigma_{dB}$ around an average value which is itself lower.

\begin{figure}
  \centerline{\includegraphics[width=\columnwidth]{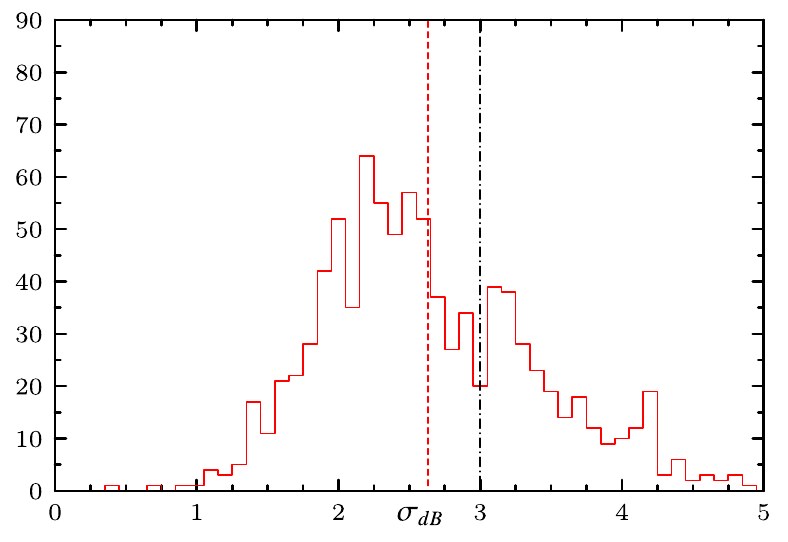}}
  \centerline{\includegraphics[width=\columnwidth]{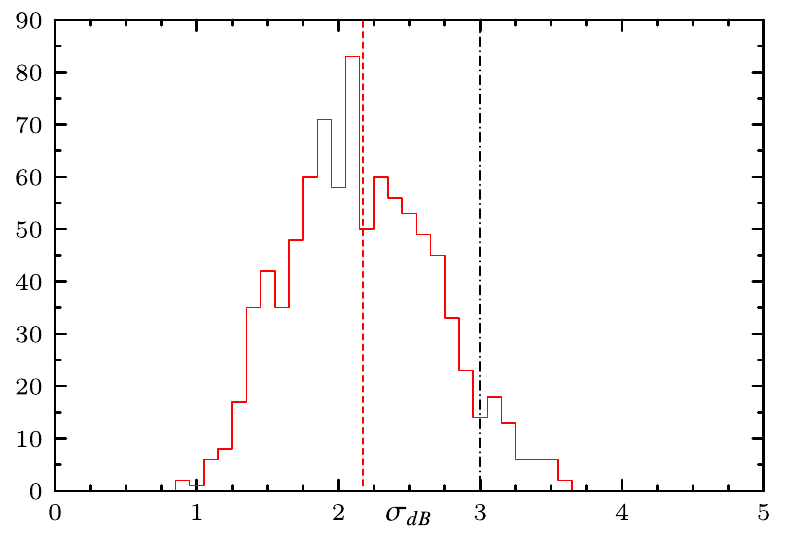}}
\caption{\label{fig:res_num}
Histograms of values of $\sigma_{dB}$ numerically obtained (via the DGT) in the two configurations of the RC shown in Fig.~\ref{fig:manip_schem}.
As previously the losses are introduced via 6 patchs of very low conductivity leading to $d\lesssim 1$.
Also shown: the limit at 3dB (black dash-dots) and the mean of $\sigma_{dB}$ (red dashes).
Above: histogram for the conventional RC with the stirrer ($\av{\sigma_{dB}}=2.63$). Below: histogram for the chaotic RC with 3 half-spheres and the stirrer ($\av{\sigma_{dB}}=2.17$).}
\end{figure}

These experimental results also confirm the validity of our numerical approach. Fig.~\ref{fig:res_num} shows the histograms of $\sigma_{dB}$, obtained through the computation of a single element of the DGT for geometries and configurations (numbers of steps of the stirrer, locations of the measurement points,...) as close as possible to those of our experiments, in particular through the introduction of Ohmic losses localized on the walls and ensuring a value of the mean modal overlap $d\lesssim 1$. The experimental and numerical behaviors agree remarkably well.

\section{Conclusions}

This paper deals with the statistics of the EM response in a reverberation chamber made chaotic by adding spherical elements in the presence of losses.
The response is modeled by taking into account the complex character of the modes of the cavity in the expression of the Dyadic Green's tensor.
A numerical study demonstrates that the criterion of uniformity of the Standard, in the vicinity of the LUF and in a regime of moderate modal overlap, is only relevant if the RC is chaotic. These numerical results are experimentally confirmed for the first time in a commercial reverberation chamber modified to be chaotic. The numerical approach and the ensuing analysis are thus validated.

\begin{acknowledgement}

The authors acknowledge the French National Research Agency (ANR) for financially supporting the project CAOREV of which this work is a part. Bernard Gay-Para is thanked for providing the artist view of Fig.~\ref{fig:manip_schem}.
\end{acknowledgement}

\bibliographystyle{IEEEtran}

\begin{thebibliography}{10}
%%---\bibitem[1]{Zhang10}
%A.-B. Zhang, C.-D. Kim, E. Yamada, F.G. Smith,
%The numerical investigation on the turbulent heat transfer,
%{\em Int. J. Heat Mass Transfer} 12: 345--365, 2010.

%\bibitem[2]{Patankar80}
%S.V. Patankar,
%{\em Numerical Heat Transfer and Fluid Flow},
%Hemisphere Publishers, New York, 1980.

\bibitem{Standard}
CISPR/A and IEC SC 77B, IEC 61000-4-21,``Electromagnetic Compatibility (EMC)- Part 4-21: Testing and Measurement Techniques - Reverberation Chamber Test Methods'', International Electrotechnical Commission (IEC) International Standard.

\bibitem{Cozza}
A. Cozza, ``The role of losses in the definition of the overmoded condition for reverberation chambers and their statistics'', {\em IEEE Trans. Electromagn. Compat.} 53: 296, 2011.
doi:10.1109/TEMC.2010.2081993

\bibitem{Hill_1}
D. Hill, ``Plane wave integral representation for fields in reverberation chambers'', {\em IEEE Trans. Electromagn. Compat.} 40: 209, 1998.
doi:10.1109/15.709418

\bibitem{Hill_2}
D. Hill, {\em Electromagnetic Fields in Cavities: Deterministic and Statistical Theories}, IEEE Press Series on Electromagnetic Wave Theory, IEEE; Wiley, 2009.

\bibitem{GROS_2013}
J.-B.~Gros, O.~Legrand, F.~Mortessagne, E.~Richalot, K.~Selemani, ``Universal behaviour of a wave chaos based electromagnetic reverberation chamber'', {\em Wave Motion} 51: 664, 2014.
doi:10.1016/j.wavemoti.2013.09.006.

\bibitem{BGS}
O.~Bohigas, M.~J.~Giannoni, and C.~Schmit.
``Characterization of chaotic quantum spectra and universality of level fluctuation laws'', {\em Phys. Rev. Lett.}, 52: 1, 1984.
doi:10.1103/PhysRevLett.52.1

\bibitem{PRGKLM}
J.-B.~Gros, U.~Kuhl, O.~Legrand, F.~Mortessagne, \emph{in preparation}.

\bibitem{KLM}
U.~Kuhl, O.~Legrand, F.~Mortessagne, ``Microwave experiments using open chaotic cavities in the realm of the effective Hamiltonian formalism'', {\em Fortschritte der Physik / Progress of Physics} 61: 404, 2013.
doi:10.1002/prop.201200101

\bibitem{GROS_2014}
J.-B.~Gros, U.~Kuhl, O.~Legrand, F.~Mortessagne, E.~Richalot, D.~V.~Savin, ``Experimental width shift distribution: a test of nonorthogonality for local and global perturbations'', preprint arXiv:1408.6472, 2014.

\bibitem{ERICSCHRO}
O.~Legrand, F.~Mortessagne, D.~Sornette, ``Spectral Rigidity in the Large Modal Overlap Regime: Beyond the Ericson-Schroeder Hypothesis'', {\em Journal de Physique I}, 5: 1003, 1995.
doi:10.1051/jp1:1995179;
Erratum: \emph{Ibid} 5: 1517, 1995.
doi:10.1051/jp1:1995214

\bibitem{KIM}
Y.-H.~Kim, U.~Kuhl, H.-J.~St\"oeckmann, P.~Brouwer, ``Measurement of Long-Range Wave-Function Correlations in an Open Microwave Billiard'', {\em Phys. Rev. Lett.}, 94: 036804, 2005.
doi:10.1103/PhysRevLett.94.036804

\bibitem{Pnini}
R. Pnini and B. Shapiro, ``Intensity fluctuations in closed and open systems'', {\em Phys. Rev. E}, 54: R1032, 1996.
doi:10.1103/PhysRevE.54.R1032

\bibitem{Arnaut}
L.~R.~Arnaut, ``Mode-stirred reverberation chambers: A paradigm for spatio-temporal complexity in dynamic electromagnetic environments'', {\em Wave Motion} 51: 673, 2014.
doi:10.1016/j.wavemoti.2013.08.007
\end{thebibliography}

\end{document}